\documentclass[10pt,final,conference,letterpaper]{IEEEtran}
\usepackage[noadjust]{cite}
\usepackage{amsthm}
\usepackage{amssymb}
\usepackage{amsmath}
\usepackage{graphicx}

\title{A Projection Method for Derivation of Non-Shannon-Type Information Inequalities}

\author{
\IEEEauthorblockN{Weidong Xu}
\IEEEauthorblockA{
Shanghai Key Lab.\\
Digital Media Proc. and Trans.\\
Dept. of Electrical Eng.\\
Shanghai Jiao Tong Univ.\\
Shanghai, 200240, China\\
Email: weidongxu@sjtu.edu.cn
}
\and
\IEEEauthorblockN{Jia Wang}
\IEEEauthorblockA{
Shanghai Key Lab.\\
Digital Media Proc. and Trans.\\
Dept. of Electrical Eng.\\
Shanghai Jiao Tong Univ.\\
Shanghai, 200240, China\\
Email: jiawang@sjtu.edu.cn
}
\and
\IEEEauthorblockN{Jun Sun}
\IEEEauthorblockA{
Shanghai Key Lab.\\
Digital Media Proc. and Trans.\\
Dept. of Electrical Eng.\\
Shanghai Jiao Tong Univ.\\
Shanghai, 200240, China\\
Email: sunjun@cdtv.org.cn
}
}

\begin{document}
\maketitle

\newtheorem{defi}{Definition}
\newtheorem{thm}{Theorem}
\newtheorem{lem}{Lemma}
\newtheorem{pro}{Proposition}

\begin{abstract}
In 1998, Zhang and Yeung found the first unconditional non-Shannon-type information inequality.
Recently, Dougherty, Freiling and Zeger gave six new unconditional non-Shannon-type information inequalities.
This work generalizes their work and provides a method to systematically derive non-Shannon-type information inequalities.
An application of this method reveals new 4-variable non-Shannon-type information inequalities.
\end{abstract}

\section{Introduction}
\label{sec_intr}

Let $n$ be a positive integer, $N$ the set $\{1,2,\cdots,n\}$, and $\mathcal{P}(N)$ its power set.
Let $\xi=(\xi_i)_{i \in N}$ be an $n$-dimension discrete random vector.
Let $\xi_I=(\xi_i)_{i \in I}$, $\forall {I \subseteq N}$.
For convenience, let $\xi_\emptyset$ be a random variable taking a fixed value with probability 1.
The entropy function $H_{\xi}$ of $\xi=(\xi_i)_{i \in N}$ maps sets $I \subseteq N$ to the Shannon entropies $H(\xi_I)$, which take value on $[0,+\infty]$.
An entropy function which takes finite values can be considered as a vector in the Euclidean space $\mathbb{R}^{\mathcal{P}(N)}$.
Let $\boldsymbol{H}_N^\textrm{ent}$ be the set of all such entropy functions.
Therefore $\boldsymbol{H}_N^\textrm{ent}$ can be viewed as a region in $\mathbb{R}^{\mathcal{P}(N)}$.
It is known that the closure of $\boldsymbol{H}_N^\textrm{ent}$, $cl(\boldsymbol{H}_N^\textrm{ent})$, is a convex cone \cite{ZY97}.
A real function $f$ on $\mathbb{R}^{\mathcal{P}(N)}$ is an information inequality if and only if $f(x) \geq 0$ for all $x \in \boldsymbol{H}_N^\textrm{ent}$.

By Shannon-type information inequalities, entropy function has the following properties:
$H_{\xi}$ is normalized, $H_{\xi}(\emptyset)=0$;
nondecreasing, $H_{\xi}(I) \leq H_{\xi}(J)$ for $I \subseteq J \subseteq N$;
submodular, $H_{\xi}(I)+H_{\xi}(J) \geq H_{\xi}(I \cup J)+H_{\xi}(I \cap J)$ for $I,J \subseteq N$.
Let $\boldsymbol{H}_N$ be the set of vectors in space $\mathbb{R}^{\mathcal{P}(N)}$ satisfying the above three properties.
Clearly, $\boldsymbol{H}_N$ is a polyhedral cone.

Obviously, $\boldsymbol{H}_N^\textrm{ent} \subseteq \boldsymbol{H}_N$.
It is known that $\boldsymbol{H}_2^\textrm{ent} = \boldsymbol{H}_2$, $cl(\boldsymbol{H}_3^\textrm{ent}) = \boldsymbol{H}_3$, and $cl(\boldsymbol{H}_N^\textrm{ent}) \neq \boldsymbol{H}_N$ for $n \geq 4$ \cite{ZY97,ZY98}.
Therefore, $\boldsymbol{H}_N$ is an outer bound of $cl(\boldsymbol{H}_N^\textrm{ent})$.
In this work, all outer bound $\boldsymbol{H}_N^\textrm{outer}$ referred will be taken for polyhedral cone.

Since Shannon-type information inequalities can not fully characterize $cl(\boldsymbol{H}_N^\textrm{ent})$, there must exist non-Shannon-type information inequalities.
First unconditional linear non-Shannon-type information inequality was found in \cite{ZY98}.
More of these non-Shannon-type information inequalities appeared in \cite{MMRV02,Zha03,DFZ06,Mat05}.
Very recently, Mat\'{u}\v{s} \cite{Mat07} derived an infinite sequence of new 4-variable and 5-variable linear information inequalities, and used them to show that $cl(\boldsymbol{H}_N^\textrm{ent})$ is not a polyhedral cone, for all $n \geq 4$.
That is to say, $cl(\boldsymbol{H}_N^\textrm{ent})$ can not be fully characterized by a finite number of linear information inequalities.

In the process of deriving non-Shannon-type information inequalities in \cite{ZY98} and \cite{DFZ06}, the authors used similar methods.
This work generalizes their concept and method, then brings forward a method to systematically derive non-Shannon-type information inequalities.
In particular, using this method, we find new 4-variable unconditional linear non-Shannon-type information inequalities.

\section{Polyhedral Cone and Its Projection}
\label{sec_poly}

In this section, we first briefly review the definition and properties of polyhedral cone, then turn to the projection of polyhedral cone.

A polyhedral cone $P$ in Euclidean space $\mathbb{R}^n$ can be represented in two ways:
by the intersection of a finite number of closed half-spaces (an \textsl{H}-representation),
\[P = \{ x \in \mathbb{R}^n \mid Ax \geq 0 \}, \textrm{ where } A \in \mathbb{R}^{r \times n};\]
or by the nonnegative linear combination of a finite set of extreme rays (a \textsl{V}-representation),
\[P = \{ x \in \mathbb{R}^n \mid x=Ry, y \in \mathbb{R}^s, y \geq 0 \}, \textrm{ where } R \in \mathbb{R}^{n \times s}.\]
Minkowski-Weyl Theorem states that these two representations are equivalent \cite{Zie95}.

In this work, a polyhedral cone is viewed as a region in Euclidean space as well as the finite set of linear inequalities representing the closed half-spaces defining the region.
In this view, polyhedral cone $\boldsymbol{H}_N$ can be considered as the region in space $\mathbb{R}^{\mathcal{P}(N)}$ as well as a set of facet-defining Shannon-type information inequalities, namely, elemental forms of Shannon's information measures \cite{Yeu97}.

A projection of a region $P$ in $\mathbb{R}^n = \mathbb{R}^{n_1} \times \mathbb{R}^{n_2}$ onto its subspace $\mathbb{R}^{n_1}$ is
\[\pi_{x_1}(P) = \left\{ x_1 \in \mathbb{R}^{n_1} \biggm| \exists x_2 \textrm{ such that } x = \left( \begin{array}{c} x_1 \\ x_2 \end{array} \right) \in P \right\}.\]
That is to say, $\pi_{x_1}$ is a projection determined by its range $\mathbb{R}^{n_1}$ and null space $\mathbb{R}^{n_2}$.
And $\pi_{x_1}(P)$ is the set of all the vectors $\pi_{x_1}(x), x \in P$.

The projection of a polyhedral cone is still a polyhedral cone.
In its \textsl{V}-representation, projection of $P$ onto $\mathbb{R}^{n_1}$ is simply
\[\pi_{x_1}(P) = \{ x_1 \in \mathbb{R}^{n_1} \mid x_1 = R_1 y, y \in \mathbb{R}^s, y \geq 0 \},\]
where $R_1 \in \mathbb{R}^{n_1 \times s}, R_2 \in \mathbb{R}^{n_2 \times s}, \left( \begin{array}{c} R_1 \\ R_2 \end{array} \right)=R$.
The projection of a polyhedral cone in its \textsl{H}-representation is more difficult.
We will detail it later in Section~\ref{sec_alg}.

Now we discuss some results relating a polyhedral cone with the half-spaces and vertexes of its projection.

A homogeneous linear inequality $a x \geq 0$ can be inferred from a system of homogeneous linear inequalities $Ax \geq 0$ if and only if the result of the linear programming problem
\[
\begin{array}{lll}
\textrm{minimize} & a x & {}\\
\textrm{subject to} & A x \geq 0\\
\end{array}
\]
is greater than or equal to $0$,
if and only if there exists a vector $y \geq 0$ so that $y^T A = a$.
The latter equivalence is in fact the well-known Farkas Lemma.
Linear programming can be used either to verify the former linear optimization problem or to find a nonnegative solution of the latter linear equations.

A set of homogeneous linear inequalities is called independent, if none of its inequalities can be inferred from other inequalities in the set.

\begin{lem}
\label{lem_i-p}
A linear inequality $a x_1 \geq 0$ can be inferred from the projection of a set of linear inequalities $A_1 x_1 + A_2 x_2 \geq 0$ onto $\mathbb{R}^{n_1}$ if and only if it can be directly inferred from $A_1 x_1 + A_2 x_2 \geq 0$.
\end{lem}
The lemma results from Projection Lemma used in block elimination algorithm \cite{Kal02}, which is proved using Farkas Lemma.

\begin{lem}
\label{lem_v-p}
A vertex $x_1$ is in the projection of a set of linear inequalities $A_1 x_1 + A_2 x_2 \geq B$ if and only if the result of the linear programming problem
\[
\begin{array}{lll}
\textrm{maximize} & (B - A_1 x_1)^T y & {}\\
\textrm{subject to} & A_2^T y = 0,\\
\textrm{} & y \geq 0\\
\end{array}
\]
is less than or equal to $0$.
\end{lem}
The equivalence can be readily proved by Farkas Lemma.

Owing to the above two lemmas, with the knowledge of the extreme rays and linear inequalities of a polyhedral cone $P_1$, one can verify whether it is the projection of polyhedral cone $P$, without actually computing the projection.

\section{Theoretical Background}
\label{sec_thm}

In this section, we present the theoretical background of the work.

In $2^n$-dimension Euclidean space $\mathbb{R}^{\mathcal{P}(N)}$, a vector is written as $x^{\mathcal{P}(N)}=(x_I)_{I \in \mathcal{P}(N)}$.
In the discussion of linear information inequality, $x_\emptyset$ is fixed to be $0$.

\subsection{Projection of $\boldsymbol{H}_N$ and $\boldsymbol{H}_N^\textrm{ent}$}
\label{subsec_prjH}

\begin{defi}
\label{def_proj}
A projection of a region $P$ in $\mathbb{R}^{\mathcal{P}(N)}$ onto its subspace $\mathbb{R}^{\mathcal{P}(M)}$ is $\pi_{x^{\mathcal{P}(M)}}(P)$, where $M = \{1,2,\cdots,m\}, m < n$.
\end{defi}

This is the same as the concept of restriction in \cite{Mat05,Mat07}.

\begin{pro}
\label{pro_hent}
$\pi_{x^{\mathcal{P}(M)}}(\boldsymbol{H}_N^\textrm{ent}) = \boldsymbol{H}_M^\textrm{ent}$.
$\pi_{x^{\mathcal{P}(M)}}(cl(\boldsymbol{H}_N^\textrm{ent})) = cl(\boldsymbol{H}_M^\textrm{ent})$.
\end{pro}

The proposition can be easily proved by truncating an $n$-dimension discrete random vector.

\begin{pro}
\label{pro_h}
$\pi_{x^{\mathcal{P}(M)}}(\boldsymbol{H}_N) = \boldsymbol{H}_M$.
\end{pro}

\subsection{Procedure for Derivation of Non-Shannon-Type Information Inequalities}
\label{subsec_prc}

Following, we state the procedure for systematically deriving non-Shannon-type information inequalities.
Below, we restrict consideration to random vectors whose entropy functions take finite values.

Say, we want to find $m$-variable information inequalities.
For any $m$-dimension random vector $\xi$, we construct a corresponding $n$-dimension ($n>m$) random vector $\zeta=(\zeta_i)_{i \in N}$, so that the marginal distribution of its first $m$ random variables, namely, the probability distribution of $\zeta_M$, equals that of $\xi$.
Let $\Delta_N$ in $\mathbb{R}^{\mathcal{P}(N)}$ be the set of entropy functions of all these $n$-dimension discrete random vectors.
Observing the above construction and recalling the definition of projection, it follows that $\pi_{x^{\mathcal{P}(M)}}(\Delta_N)=\boldsymbol{H}_M^\textrm{ent}$.
Therefore, if we can to some extent characterize $\Delta_N$, thus its projection $\pi_{x^{\mathcal{P}(M)}}(\Delta_N)$, we can finally arrive at an outer bound of $\boldsymbol{H}_M^\textrm{ent}$.
Obviously, $\boldsymbol{H}_N$ is a trivial outer bound of $\Delta_N$.
Since $\pi_{x^{\mathcal{P}(M)}}(\boldsymbol{H}_N)=\boldsymbol{H}_M$, it provides no new information about $\boldsymbol{H}_M^\textrm{ent}$.
However, for properly designed $\zeta$, some better outer bound of $\Delta_N$ can be reached.

In the latter part of the section, we focus on the technique used in \cite{DFZ06} to construct the $n$-dimension random vector $\zeta$ and the outer bound of $\Delta_N$.

We restate a lemma in \cite[Lemma 14.8]{Yeu02}.

\begin{lem}
\label{lem_n2n+1}
Given an $n$-dimension random vector $\xi=(\xi_i)_{i \in N}$, there exists a random variable $\xi_{n+1}$, such that:
\begin{enumerate}
\item The joint probability distribution of $(\xi_{n+1},\xi_I)$ equals that of $(\xi_k,\xi_I)$,
\item Markov chain $\xi_{\{k\} \cup J} \rightarrow \xi_I \rightarrow \xi_{n+1}$ holds,
\end{enumerate}
where $\{k\},I, J$ are disjoint subsets of $N$.
\end{lem}

In \cite{DFZ06}, random variable $\xi_{n+1}$ is called a $\xi_J$-copy of $\xi_k$ over $\xi_I$.

Therefore, with $\xi_{n+1}$, $\zeta=(\xi_1,\cdots,\xi_n,\xi_{n+1})$ is an $n+1$-dimension random vector.
And
\[H_\zeta(\{n+1\} \cup I_1)=H_\zeta(\{k\} \cup I_1),I_1 \subseteq I,\]
\[H_\zeta(\{n+1\} \cup I)+H_\zeta(I \cup J_1)=H_\zeta(\{n+1\} \cup I \cup J_1)+H_\zeta(I),\]
\begin{flushright}
$J_1 \subseteq \{k\} \cup J,J_1 \neq \emptyset$,
\end{flushright}
for all $\zeta$.

For constructing an $n$-dimension random vector from $m$-dimension random vector $\xi$, we start with $\xi$, then add auxiliary random variables $\xi_{m+1},\cdots,\xi_n$ according to Lemma~\ref{lem_n2n+1}, choosing parameters $k, I, J$ every time.
The process ends with an $n$-dimension random vector $\zeta=(\xi_1,\cdots,\xi_m,\xi_{m+1},\cdots,\xi_n)$.
Besides all $n$-variable Shannon-type information inequalities, depending on the parameters chosen, the entropy function of $\zeta$ also satisfies the above two kinds of equalities.
Let $C_N$ denote the set of all these additional linear equalities.
The region of the entropy functions of all $\zeta$, $\Delta_N$, is outer bounded by the polyhedral cone $\boldsymbol{H}_N \cap C_N$.
Recalling that $\boldsymbol{H}_M^\textrm{ent}=\pi_{x^{\mathcal{P}(M)}}(\Delta_N)$, it follows that $\boldsymbol{H}_M^\textrm{ent} \subseteq cl(\boldsymbol{H}_M^\textrm{ent}) \subseteq \pi_{x^{\mathcal{P}(M)}}(\boldsymbol{H}_N \cap C_N)$.
The projection of $\boldsymbol{H}_N \cap C_N$ can be explicitly calculated.
If $\pi_{x^{\mathcal{P}(M)}}(\boldsymbol{H}_N \cap C_N)$ is strictly smaller than $\boldsymbol{H}_M$, we then have a better characterization of $\boldsymbol{H}_M^\textrm{ent}$.
In other words, some of the facet-defining linear inequalities of $\pi_{x^{\mathcal{P}(M)}}(\boldsymbol{H}_N \cap C_N)$ must be $m$-variable non-Shannon-type information inequalities.

In the above procedure, alternatively, we could use any known outer bound $\boldsymbol{H}_N^\textrm{outer}$ instead of $\boldsymbol{H}_N$.
That is to say, we calculate the projection of $\boldsymbol{H}_N^\textrm{outer} \cap C_N$ instead.
This means using known linear non-Shannon-type information inequalities in the derivation of new non-Shannon-type information inequalities.

We delay the application of above procedure to Section~\ref{sec_res}.

\section{Projection Algorithm}
\label{sec_alg}

Although projection of a polyhedral cone in its \textsl{V}-representation is simply the projection of all its extreme rays, in this work, we prefer the \textsl{H}-representation, partly because we are interested in linear information inequalities, and partly because the \textsl{V}-representation of an outer bound might be exponentially more complex than its \textsl{H}-representation.

In order to derive non-Shannon-type information inequalities using the theory and method mentioned in Section~\ref{sec_thm}, we must actually calculate the projection of a polyhedral cone.
However, even for deriving $4$-variable non-Shannon-type information inequalities by projecting the polyhedral cone outer bounding the entropy functions of all corresponding $6$-dimension random vectors, it requires calculating the projection of a $2^6$-dimension polyhedral cone consisting of hundreds of linear inequalities onto a $2^4$-dimension space.
Moreover, there is no explicit complexity results known for polyhedron projection \cite{Kal02}.

Classical projection methods such as Fourier-Motzkin elimination and block elimination are too computationally inefficient to project high dimension polyhedron \cite{HLL92,Kal02}.
And the degeneracy of the polyhedral cone may render other projection algorithms, such as ESP \cite{JKM04}, inefficient.

\begin{figure}[!t]
\centering
\includegraphics{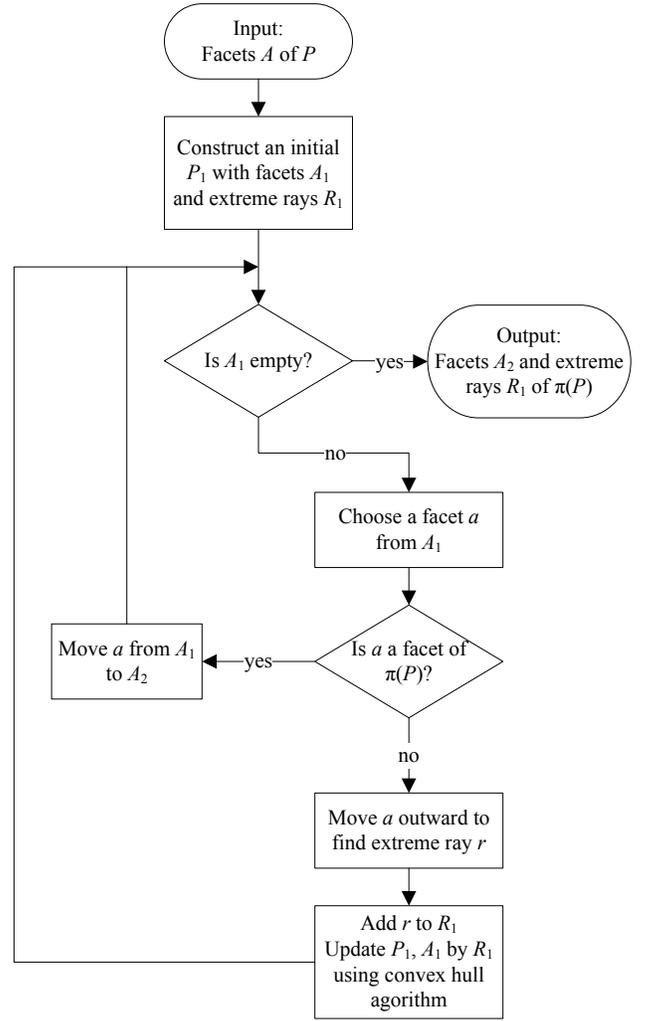}
\caption{Flowchart of convex hull method for projection}
\label{fig_fcchm}
\end{figure}

We turn to convex hull method (CHM) proposed in \cite{LL93}.
It works directly in the projection space.
The flowchart of CHM is depicted in Fig.~\ref{fig_fcchm}.
For projecting a polyhedral cone $P$ in $\mathbb{R}^n$ onto $\mathbb{R}^{n_1}$, the algorithm incrementally constructs a polyhedral cone $P_1$ in $\mathbb{R}^{n_1}$ while maintaining its double description pair $\left( \left( \begin{array}{c} A_1 \\ A_2 \end{array} \right), R_1 \right)$.
The output is $\pi_{x_1}(P) = P_1$, represented by double description pair $(A_2, R_1)$.

The convex hull algorithm used in the CHM can be implemented using any incremental algorithm.
We adopt Fourier-Motzkin elimination (the dual of double description method \cite{FP96}), since it appears to deal well with degeneracy \cite{ABS97}.

We can take advantage of the incremental nature of CHM to further reduce its computational complexity.
Let $\boldsymbol{H}_M^\textrm{inner}$ be an inner bound of $cl(\boldsymbol{H}_M^\textrm{ent})$.
Then the extreme rays of $\boldsymbol{H}_M$ in $\boldsymbol{H}_M^\textrm{inner}$ must be the extreme rays of $cl(\boldsymbol{H}_M^\textrm{ent})$, as well as the extreme rays of any $\pi_{x^{\mathcal{P}(M)}}(\boldsymbol{H}_N^\textrm{outer} \cap C_N)$.
If the polyhedral cone generated by these extreme rays is full dimensional (except on axis $x_\emptyset$), we can then skip the initialization and immediately start CHM from this approximation.
This method can be used for projection onto $\mathbb{R}^{\mathcal{P}(\{1,2,3,4\})}$, since the inner bound of $cl(\boldsymbol{H}_4^\textrm{ent})$ is known \cite{MS95,ZY98}.

\section{New $4$-variable Non-Shannon-Type Information Inequalities}
\label{sec_res}

In this section, we focus on $4$-variable non-Shannon-type information inequalities.
They are of theoretical importance, meanwhile, low dimension polyhedral cones are more computationally tractable.
In the following, we discuss two different approaches for calculating new non-Shannon-type information inequalities using the procedure introduced in Section~\ref{subsec_prc}.
Then the results are used to help derive an infinite sequence of $4$-variable non-Shannon-type information inequalities.

\subsection{Projection by Adding More Random Variables}
\label{subsec_addrv}

First we add the 5th and 6th random variables, as what is shown in the proof of \cite[Theorem III.1]{DFZ06}.
For any probability distribution of $4$-dimension random vector $\xi$, by Lemma~\ref{lem_n2n+1}, let $\xi_5$ be a $\xi_4$-copy of $\xi_3$ over $\xi_{\{1,2\}}$, $\xi_6$ be a $\xi_2$-copy of $\xi_3$ over $\xi_{\{1,4,5\}}$.
This results in a $6$-dimension random vector $\zeta=(\xi_1,\cdots,\xi_6)$.
And the corresponding polyhedral cone $C_6$ has 18 equalities.
$\pi_{x^{\mathcal{P}(\{1,2,3,4\})}}(\boldsymbol{H}_6 \cap C_6)$ is a polyhedral cone defined by 35 linear information inequalities, which reveals Zhang-Yeung inequality and the third and fifth Dougherty-Freiling-Zeger inequalities.

We use more than $6$ random variables.
For instance, let $\xi_7$ be a $\xi_4$-copy of $\xi_2$ over $\xi_{\{1,3,5,6\}}$.
And $\zeta=(\xi_1,\cdots,\xi_7)$.
The corresponding polyhedral cone $C_7$ has 19 more equalities.
The resulting projection $\pi_{x^{\mathcal{P}(\{1,2,3,4\})}}(\boldsymbol{H}_7 \cap C_7)$ is a polyhedral cone defined by 56 linear information inequalities, which reveals 13 independent non-Shannon-type information inequalities unknown before.
Due to space limitations we just list several of them.
\begin{IEEEeqnarray*}{l}
{-}\:56H(\xi_{\{1\}})-4H(\xi_{\{2\}})-19H(\xi_{\{3\}})\\
{+}\:45H(\xi_{\{1,2\}})+67H(\xi_{\{1,3\}})+22H(\xi_{\{2,3\}})\\
{+}\:23H(\xi_{\{1,4\}})-8H(\xi_{\{2,4\}})+9H(\xi_{\{3,4\}})\\
{-}\:55H(\xi_{\{1,2,3\}})-24H(\xi_{\{1,3,4\}}) \geq 0,
\end{IEEEeqnarray*}
\begin{IEEEeqnarray*}{l}
{-}\:34H(\xi_{\{1\}})-2H(\xi_{\{2\}})-11H(\xi_{\{3\}})-H(\xi_{\{4\}})\\
{+}\:27H(\xi_{\{1,2\}})+40H(\xi_{\{1,3\}})+12H(\xi_{\{2,3\}})\\
{+}\:15H(\xi_{\{1,4\}})-5H(\xi_{\{2,4\}})+7H(\xi_{\{3,4\}})\\
{-}\:32H(\xi_{\{1,2,3\}})-16H(\xi_{\{1,3,4\}}) \geq 0,
\end{IEEEeqnarray*}
\begin{IEEEeqnarray*}{l}
{-}\:28H(\xi_{\{1\}})-H(\xi_{\{2\}})-10H(\xi_{\{3\}})-2H(\xi_{\{4\}})\\
{+}\:22H(\xi_{\{1,2\}})+34H(\xi_{\{1,3\}})+11H(\xi_{\{2,3\}})\\
{+}\:13H(\xi_{\{1,4\}})-4H(\xi_{\{2,4\}})+6H(\xi_{\{3,4\}})\\
{-}\:28H(\xi_{\{1,2,3\}})-13H(\xi_{\{1,3,4\}}) \geq 0.
\end{IEEEeqnarray*}
The novelty and independence of these information inequalities can be verified by finding an extreme ray which does not satisfy one of the inequalities, but satisfies all $4$-variable Shannon-type information inequalities together with all substituted forms of Zhang-Yeung inequality, Dougherty-Freiling-Zeger inequalities, Mat\'{u}\v{s} inequalities, and the rest of these inequalities.
The verification can be done by linear programming.

Nevertheless, without calculation (projection or verification), we do not know beforehand how these auxiliary random variables would affect the resulting projection.
This hinders us from systematically designing the random vector $\zeta$.

\subsection{Projection Using Improved Outer Bound}
\label{subsec_outbnd}

Alternatively, we choose not to add that many auxiliary random variables, but to use an outer bound strictly smaller than $\boldsymbol{H}_N$.

Let $\xi_5$ be a $\xi_4$-copy of $\xi_3$ over $\xi_{\{1,2\}}$.
During process in Section~\ref{subsec_addrv}, it is already known that $\pi_{x^{\mathcal{P}(\{1,2,3,4\})}}(\boldsymbol{H}_5 \cap C_5)$ only reveals Zhang-Yeung non-Shannon-type information inequality,
\begin{IEEEeqnarray}{l}
\label{eq:nii0}
{-}\:2H(\xi_{\{1\}})-2H(\xi_{\{2\}})-H(\xi_{\{3\}}) \nonumber\\
{+}\:3H(\xi_{\{1,2\}})+3H(\xi_{\{1,3\}})+3H(\xi_{\{2,3\}}) \nonumber\\
{+}\:H(\xi_{\{1,4\}})+H(\xi_{\{2,4\}})-H(\xi_{\{3,4\}}) \nonumber\\
{-}\:4H(\xi_{\{1,2,3\}})-H(\xi_{\{1,2,4\}}) \geq 0.
\end{IEEEeqnarray}
An outer bound $\boldsymbol{H}_4^{\textrm{outer}(1)}$ is constructed from $\boldsymbol{H}_4$ and Zhang-Yeung inequality.

Through substitution, an outer bound $\boldsymbol{H}_5^{\textrm{outer}(1)}$ can be constructed from $\boldsymbol{H}_4^{\textrm{outer}(1)}$.
We repeat the procedure, but project $\boldsymbol{H}_5^{\textrm{outer}(1)} \cap C_5$ rather than $\boldsymbol{H}_5 \cap C_5$.
The resulting $\pi_{x^{\mathcal{P}(\{1,2,3,4\})}}(\boldsymbol{H}_5^{\textrm{outer}(1)} \cap C_5)$ is a polyhedral cone defined by 55 linear information inequalities, 6 of which are independent information inequalities that can not be inferred from $\boldsymbol{H}_4^{\textrm{outer}(1)}$.
Among them, the following is unknown before,
\begin{IEEEeqnarray}{l}
\label{eq:nii1}
{-}\:10H(\xi_{\{1\}})-10H(\xi_{\{2\}})-H(\xi_{\{3\}}) \nonumber\\
{+}\:17H(\xi_{\{1,2\}})+10H(\xi_{\{1,3\}})+10H(\xi_{\{2,3\}}) \nonumber\\
{+}\:4H(\xi_{\{1,4\}})+4H(\xi_{\{2,4\}})-3H(\xi_{\{3,4\}}) \nonumber\\
{-}\:16H(\xi_{\{1,2,3\}})-5H(\xi_{\{1,2,4\}}) \geq 0.
\end{IEEEeqnarray}
Outer bound $\boldsymbol{H}_4^{\textrm{outer}(2)}$ is constructed from $\boldsymbol{H}_4^{\textrm{outer}(1)}$ together with these new information inequalities.

It can be observed that the above procedure is indeed a function which maps $\boldsymbol{H}_4$ to $\boldsymbol{H}_4^{\textrm{outer}(1)}$, and subsequently $\boldsymbol{H}_4^{\textrm{outer}(1)}$ to $\boldsymbol{H}_4^{\textrm{outer}(2)}$.

\subsection{Infinite Sequence of $4$-variable Non-Shannon-Type Information Inequalities}
\label{subsec_infseq}

Though the projection algorithm will only reveal finite number of non-Shannon-type linear information inequalities, the information acquired from the results may still help shed some light on the infinite sequence of linear information inequalities.

Once a new $m$-variable linear information inequality is derived by projecting some $\boldsymbol{H}_N^\textrm{outer} \cap C_N$, it can be written as a nonnegative linear combination of those $n$-variable information inequalities and additional equalities, as discussed in Section~\ref{sec_poly}.
This nonnegative linear combination can be viewed as an explicit proof of this newly derived information inequality.

It can be observed that Zhang-Yeung information inequality (\ref{eq:nii0}) and information inequality (\ref{eq:nii1}) bear some similarity.
Moreover, the explicit proofs corresponding to the two information inequalities share a similar structure.
Therefore, with several more iterations and some guesswork, we derive the following infinite sequence of $4$-variable non-Shannon-type information inequalities.
\begin{IEEEeqnarray}{l}
\label{eq:isii1}
\left( 2^{s-1} - \frac{\sqrt2}{4} S_+ + \frac{\sqrt2}{4} S_- \right) \left[ H(\xi_{\{1\}}) + H(\xi_{\{2\}}) \right] \nonumber\\
{-}\: H(\xi_{\{3\}}) + \left( 1 - 3 \cdot 2^{s-1} + \frac{\sqrt2}{2} S_+ - \frac{\sqrt2}{2} S_- \right) H(\xi_{\{1,2\}}) \nonumber\\
{+}\: \left( \frac{1}{4} S_+ + \frac{1}{4} S_- \right) \left[ H(\xi_{\{1,3\}}) +  H(\xi_{\{2,3\}}) \right] \nonumber\\
{+}\: \left( \frac{\sqrt2-1}{4} S_+ - \frac{\sqrt2+1}{4} S_- \right) \left[ H(\xi_{\{1,4\}}) + H(\xi_{\{2,4\}}) \right] \nonumber\\
{+}\: \left( 1 - 2^{s-1} \right) H(\xi_{\{3,4\}}) + \left( 2^{s-1} - \frac{1}{2} S_+ - \frac{1}{2} S_- \right) H(\xi_{\{1,2,3\}}) \nonumber\\
{-}\: \left( 1 - 2^{s-1} + \frac{\sqrt2-1}{2} S_+ - \frac{\sqrt2+1}{2} S_- \right) H(\xi_{\{1,2,4\}}) \geq 0, \nonumber\\
\textrm{where } S_+ = \left(2+\sqrt2\right)^s, S_-=\left(2-\sqrt2\right)^s, \textrm{ and } s \in \mathbb{N}.
\end{IEEEeqnarray}
A rigorous proof can be easily obtained through mathematical induction.

When $s=1$, information inequality (\ref{eq:isii1}) corresponds to a Shannon-type information inequality.
For $s=2$ and $3$, it is Zhang-Yeung inequality and information inequality (\ref{eq:nii1}), respectively.

\enlargethispage{-0.3in}

\section{Discussion and Future Work}
\label{sec_dis}

The concept and usage of projection is not limited to the theory and application mentioned above.
The notion of inference rule in \cite{MMRV02} and inner adhesivity in \cite{Mat07} and their methods for deriving non-Shannon-type information inequalities can also be regarded as a practice of projection.
For example, in the proof of the 5-variable non-Shannon-type information inequality in \cite{MMRV02}, the application of the inference rule is equivalently the following set of additional equalities,
\[x_{\{3,4,5\}} + x_{\{3,4\} \cup J} - x_{\{3,4,5\} \cup J} - x_{\{3,4\}} = 0,\]
\begin{flushright}
$J \subseteq \{1,2\}, J \neq \emptyset$,
\end{flushright}
and the requirement that no term in the resulting inequality has any of ${\{1,2\}}$ together with $\{5\}$ specifies the range of the projection, i.e. the subspace consisting of all vectors with $x_{I \cup J \cup \{5\}}=0, I \subseteq \{3,4\}, J \subseteq \{1,2\}, J \neq \emptyset$.

The procedure used in Section~\ref{subsec_addrv} can be generalized and its limit property is to be investigated.
Let $\boldsymbol{H}_{M \downarrow N}^\textrm{outer}$ denote the intersection of all projections $\pi_{x^{\mathcal{P}(M)}}(\boldsymbol{H}_N \cap C_N)$ derived using the procedure described in Section~\ref{subsec_prc} applying Lemma~\ref{lem_n2n+1}.
For example, it can be shown that $\boldsymbol{H}_{4 \downarrow 5}^\textrm{outer}$ is in fact the region enclosed by all $4$-variable Shannon-type information inequalities and all substituted forms of Zhang-Yeung inequality, and that $\boldsymbol{H}_{4 \downarrow 6}^\textrm{outer}$ is the region enclosed by all $4$-variable Shannon-type information inequalities and all substituted forms of Zhang-Yeung inequality and Dougherty-Freiling-Zeger inequalities.
It is easy to see that $\boldsymbol{H}_{M \downarrow N}^\textrm{outer} \subseteq \boldsymbol{H}_{M \downarrow N'}^\textrm{outer}$, for $n > n'$.
Therefore, we define $\boldsymbol{H}_{M \downarrow \mathbb{N}}^\textrm{outer} = \bigcap_{n > m} \boldsymbol{H}_{M \downarrow N}^\textrm{outer}$.
And obviously $\boldsymbol{H}_{M \downarrow \mathbb{N}}^\textrm{outer} \supseteq \boldsymbol{H}_M^\textrm{ent}$.
Then, the critical issue is whether $\boldsymbol{H}_{M \downarrow \mathbb{N}}^\textrm{outer} = cl(\boldsymbol{H}_M^\textrm{ent})$.
If the above equation holds, then this work presents an explicit characterization of $cl(\boldsymbol{H}_M^\textrm{ent})$, and we are capable of calculating and approximating it.
Otherwise, the problem is whether there exist other methods of adding random variables and other kinds of additional inequalities that can be used as $C_N$, so that after a similar process the newly obtained $\boldsymbol{H}_{M \downarrow \mathbb{N}}^\textrm{outer}$ would coincide with $cl(\boldsymbol{H}_M^\textrm{ent})$.

Similarly, the procedure described in Section~\ref{subsec_outbnd} can be generalized.
Let $\sigma$ be a function defined on the set of polyhedral cones in $\mathbb{R}^{\mathcal{P}(M)}$, which maps an outer bound $\boldsymbol{H}_M^\textrm{outer}$ to the intersection of all projections $\pi_{x^{\mathcal{P}(M)}}(\boldsymbol{H}_N^\textrm{outer} \cap C_N)$, for fixed $n$.
Clearly, $\sigma^{k+1}(\boldsymbol{H}_M) \subseteq \sigma^k(\boldsymbol{H}_M)$, and $\sigma^k(\boldsymbol{H}_M) \supseteq \boldsymbol{H}_M^\textrm{ent}$.
Then, the problem is whether $\sigma^{k+1}(\boldsymbol{H}_M)$ is strictly smaller than $\sigma^k(\boldsymbol{H}_M)$, and whether $\lim_{k \to \infty} \sigma^k(\boldsymbol{H}_M) = cl(\boldsymbol{H}_M^\textrm{ent})$.

These call for further investigation.


\begin{thebibliography}{11}
\providecommand{\url}[1]{#1}
\csname url@samestyle\endcsname
\providecommand{\newblock}{\relax}
\providecommand{\bibinfo}[2]{#2}
\providecommand{\BIBentrySTDinterwordspacing}{\spaceskip=0pt\relax}
\providecommand{\BIBentryALTinterwordstretchfactor}{4}
\providecommand{\BIBentryALTinterwordspacing}{\spaceskip=\fontdimen2\font plus
\BIBentryALTinterwordstretchfactor\fontdimen3\font minus\fontdimen4\font\relax}
\providecommand{\BIBforeignlanguage}[2]{{%
\expandafter\ifx\csname l@#1\endcsname\relax
\typeout{** WARNING: IEEEtran.bst: No hyphenation pattern has been}%
\typeout{** loaded for the language `#1'. Using the pattern for}%
\typeout{** the default language instead.}%
\else
\language=\csname l@#1\endcsname
\fi
#2}}
\providecommand{\BIBdecl}{\relax}
\BIBdecl

\bibitem{ZY97} Z.~Zhang and R.~W.~Yeung, ``A non-Shannon-type conditional inequality of information quantities,'' \emph{IEEE Trans. Inform. Theory}, vol.~43, no.~6, pp.~1982--1985, Nov.~1997.
\bibitem{ZY98} Z.~Zhang and R.~W.~Yeung, ``On characterization of entropy function via information inequalities,'' \emph{IEEE Trans. Inform. Theory}, vol.~44, no.~4, pp.~1440--1452, Jul.~1998.
\bibitem{MMRV02} K.~Makarychev, Y.~Makarychev, A.~Romashchenko, and N.~Vereshchagin, ``A new class of non-Shannon-type inequalities for entropies,'' \emph{Commun. Inf. and Syst.}, vol.~2, no.~2, pp.~147--166, Dec.~2002.
\bibitem{Zha03} Z.~Zhang, ``On a new non-Shannon type information inequality,'' \emph{Commun. Inf. and Syst.}, vol.~3, no.~1, pp.~47--60, June~2003.
\bibitem{DFZ06} R.~Dougherty, C.~Freiling, and K.~Zeger, ``Six new non-Shannon information inequalities,'' in \emph{Proc. IEEE Int. Symp. Information Theory}, Seattle, WA, Jul.~2006, pp.~233--236.
\bibitem{Mat05} F.~Mat\'{u}\v{s}, ``Inequalities for Shannon entropies and adhesivity of polymatroids,'' in \emph{Proc. 9th Canadian Workshop on Information Theory}, McGill University, Montr\'{e}al, QC, Canada, 2005, pp.~28--31.
\bibitem{MS95} F.~Mat\'{u}\v{s} and M.~Studen\'{y}, ``Conditional independences among four random variables I,'' \emph{Combinatorics, Probability and Computing}, vol.~4, no.~4, pp.~269--278, 1995.
\bibitem{Yeu97} R.~W.~Yeung, ``A framework for linear information inequalities,'' \emph{IEEE Trans. Inform. Theory}, vol.~43, no.~6, pp.~1924--1934, Nov.~1997.
\bibitem{Yeu02} R.~W.~Yeung, \emph{A First Course in Information Theory}. New York: Kluwer Academic/Plenum, 2002.
\bibitem{Zie95} G.~M.~Ziegler, \emph{Lectures on Polytopes}. Springer-Verlag, 1995.
\bibitem{HLL92} T.~Huynh, C.~Lassez, and J-L.~Lassez, ``Practical issues on the projection of polyhedral sets,'' \emph{Annals of Mathematics and Artificial Intelligence}, vol.~6, pp.~295--315, Nov.~1992. 
\bibitem{Kal02} B.~Kaluzny, ``Polyhedral computation: a survey of projection methods,'' Class Project, 2002.
\bibitem{LL93} C.~Lassez and J-L.~Lassez, ``Quantifier elimination for conjunctions of linear constraints via a convex hull algorithm,'' \emph{Symbolic and Numerical Computation for Artificial Intelligence}, Donald, Kapur, and Mundy Ed., Academic Press, 1993.
\bibitem{FP96} K.~Fukuda and A.~Prodon, ``Double description method revisited,'' \emph{Combinatorics and Computer Science}, vol.~1120 of LNCS. Springer-Verlag, pp.~91--111, 1996.
\bibitem{JKM04} C.~N.~Jones, E.~C.~Kerrigan, and J.~M.~Maciejowski, ``Equality Set Projection: A new algorithm for the projection of polytopes in halfspace representation,'' Technical Report CUED/F-INFENG/TR.463, 2004.
\bibitem{ABS97} D.~Avis, D.~Bremner, and R.~Seidel, ``How good are convex hull algorithms?'' \emph{Computational Geometry}, vol.~7, no.~5, pp.~265--301, April~1997.
\bibitem{Mat07} F.~Mat\'{u}\v{s}, ``Infinitely many information inequalities,'' in \emph{Proc. IEEE Int. Symp. Information Theory}, Nice, France, June~2007, pp.~41--44.
\end{thebibliography}
\end{document}